# A Comparison of the Performance of the Molecular Dynamics Simulation Package GROMACS Implemented in the SYCL and CUDA Programming Models

Leonard Apanasevich, Yogesh Kale, Himanshu Sharma, and Ana Marija Sokovic
University of Illinois Chicago,
USA

*Abstract*—For many years, systems running Nvidia-based GPU architectures have dominated the heterogeneous supercomputer landscape. However, recently GPU chipsets manufactured by Intel and AMD have cut into this market and can now be found in some of the world's fastest supercomputers. The June 2023 edition of the TOP500 list of supercomputers ranks the Frontier supercomputer at the Oak Ridge National Laboratory in Tennessee as the top system in the world. This system features AMD Instinct 250X GPUs and is currently the only true exascale computer in the world. In the near future, another exascale system, Aurora, equipped with Intel Delta GPUs, is expected to come online at Argonne National Laboratory in Illinois.

As the use of different GPU architectures becomes more prevalent in today's supercomputing centers, it is becoming crucial to have a programming model that could support different platforms without the need for separate codebases (Pascuzzi, 2021). The first framework that enabled support for heterogeneous platforms across multiple hardware vendors was OpenCL, in 2009. Since then a number of frameworks have been developed to support vendor-agnostic heterogeneous environments including OpenMP, OpenCL, Kokkos, and SYCL. SYCL, which combines the concepts of OpenCL with the flexibility of single-source C++, is one of the more promising programming models for heterogeneous computing devices. One key advantage of this framework is that it provides a higher-level programming interface that abstracts away many of the hardware details than the other frameworks. This makes SYCL easier to learn and to maintain across multiple architectures and vendors.

In recent years, there has been growing interest in using heterogeneous computing architectures to accelerate molecular dynamics simulations. Some of the more popular molecular dynamics simulations include Amber, NAMD, and Gromacs. However, to the best of our knowledge, only Gromacs has been successfully ported to SYCL to date. In this paper, we compare the performance of GROMACS compiled using the SYCL and CUDA frameworks for a variety of standard GROMACS benchmarks. In addition, we compare its performance across three different Nvidia GPU chipsets, P100, V100, and A100.

## I. Introduction

Accelerators, such as GPUs, FPGAs (Field-Programmable Gate Arrays), and ASICs (Application-Specific Integrated Circuits), have become essential components in various domains, including artificial intelligence, scientific computing, and embedded systems. These accelerators are designed to enhance the performance and efficiency of specific tasks by offloading computation from general-purpose processors. GPUs in particular have been widely used as accelerators in many deep learning and artificial intelligence applications due to their highly parallel architecture and ability to perform massive computations in parallel [1]. Introduced in 2006 by NVIDIA, CUDA has emerged as the reigning framework for developing code specifically designed to run on GPUs. CUDA offers a rich assortment of libraries that serve an ever-growing number of domains and applications. It is important to note that CUDA-enabled applications are often able to achieve exceptional levels of performance and efficiency on NVIDIA hardware because CUDA is specifically designed to run on NVIDIA GPUs, which allows developers to leverage the underlying hardware at a low level. However, this decision also makes it impossible to port CUDA to GPUs manufactured by other vendors. SYCL was introduced as a new cross-platform abstraction layer to provide an efficient way for single-source heterogeneous computing using C++ (Khronos) in March 2014. SYCL allows developers to write one code that can be executed on various hardware accelerators, such as GPUs, FPGAs, and ASICs, enabling efficient utilization of these resources [2].

## II. Previous Work

There have been a number of recent studies that attempt to compare the performance of SYCL and CUDA, each typically focusing on a particular application or library. Overall, these papers suggest that SYCL provides better portability and a higher level of abstraction than CUDA, but that CUDA may still have an advantage in terms of performance optimization for NVIDIA GPUs. However, the specific performance characteristics may depend on the particular benchmark or application being used. Below is a brief overview of some recent comparisons found in the literature. In [3], the authors compare the performance of SYCL and CUDA on Tesla V100 GPUs for three different benchmark applications: BabelSTREAM [4], which measures memory transfer rates to and from global device memory; Mixbench [5], which evaluates the execution limits for GPUs on mixed operational intensity kernels; and a custom built Tiled Matrix-Multiplication application, which benchmarked various matrix operations for two different memory management approaches. The authors found that both SYCL and CUDA provide similar performance for most benchmarks, with CUDA generally providing slightly better performance for large matrix sizes. They also observed that many of the performance differences could be ascribed to matrix ordering and choices of how to load memory. Reference ref used the RAJA Performance Suite [ref] to compare execution performance for SYCL and CUDA against various loop-based computational kernels. This study used the hipSYCL toolchain [3] to compile SYCL kernels directly into CUDA code and found that in most cases the SYCL kernels to be competitive with native CUDA kernels. The performance of Fast Fourier Transform (FFT) libraries developed for SYCL versus FFT libraries developed for Nvidia was studied in [6] found the SYCL libraries typically ran ~30% slower than their CUDA counterparts, after differences in kernel launch overheads were taken into account. In[7], the performance and code portability of ADEPT, a widely used bioinformatics sequence alignment kernel, has been evaluated across various vendor GPUs and programming models and found that the performance portability across different GPU architectures was rather poor and concluded that the ADEPT developers has been tuned and optimized the program for NVIDIA GPUs. Overall, the choice between CUDA and SYCL depends on the specific needs of the application and the target hardware. CUDA is a good choice for applications that require high performance and low-level access to the GPU hardware, while SYCL is a better choice for applications that need to be portable across different architectures and vendors.

## III. Introduction to Gromacs

GROMACS (GROningen MAchine for Chemical Simulations) [8] is a widely used software package for biomolecular simulations. It has been optimized for high-performance computing environments and has been employed in various studies, including simulating aptamer-peptide binding in biosensor applications and implementing dimer metadynamics. In recent years, there has been a significant focus on leveraging the computational power of graphics processing units (GPUs) to accelerate GROMACS simulations. The integration of GPUs in GROMACS has revolutionized the field of molecular dynamics (MD) simulations, leading to significant improvements in terms of performance and scalability. GPU-accelerated simulations have demonstrated speedups of up to 10 times compared to CPU-only simulations [9]. GPU-accelerated GROMACS simulations have been utilized to study protein folding dynamics, protein-ligand binding, and protein unfolding under various conditions ([10], [11], [12]). These simulations provide valuable insights into the behavior and function of biomolecules, aiding in drug discovery and understanding biological processes. GROMACS has supported GPU acceleration since version 4.5 and natively integrated GPU support since version 4.6 [9]. The native GPU support in GROMACS combines reformulated molecular dynamics algorithms with a heterogeneous parallelization scheme that utilizes both multicore CPUs and GPU accelerators [9]. Initially developed using CUDA, GROMACS shifted its focus to creating a portable multi-vendor backend, utilizing OpenCL as a standards-based GPU API [9]. Most recently, the GROMACS codebase has been updated to include support for SYCL.

GROMACS offers a range of functionalities to accommodate various types of force calculations. In terms of computational expense, the three most significant classes for most simulations are as follows:

- Non-bonded short-range forces: These forces account for direct interactions between particles within a specified cutoff distance. Only particles within this range are considered to interact directly with each other.
- Particle Mesh Ewald (PME) long-range forces: To model forces over larger distances, GROMACS employs the PME method. This approach utilizes Fourier transforms to perform calculations in Fourier space, significantly reducing the computational cost compared to direct calculations in real space.
- Bonded short-range forces: GROMACS also incorporates bonded short-range forces to capture specific behaviors of bonded particles. For example, when two covalently bonded atoms are stretched, a harmonic potential is applied to account for the bond's characteristics. In summary, GROMACS provides functionality to handle non-bonded short-range forces, PME long-range forces, and bonded short-range forces, which collectively contribute to accurate and efficient force calculations in simulations. The solution in the GROMACS 2020 version is full GPU enablement of these key computational sections.

## IV. Performance Analysis

The performance of Gromacs in the CUDA and SYCL frameworks was measured using the benchmark datasets found in [13] and [14]. These datasets have been carefully selected by MD experts, ensuring the availability of high-quality ground truth data for accurate benchmarking comparisons. The GROMACS version employed for these comparisons was 2023.1, and all experiments were executed on High-Performance Computing (HPC) systems utilizing the Apptainer/Singularity containerization platform.

In total we compare CUDA and SYCL implementations of Gromacs for 4 different datasets: benchMEM, and three water datasets. The benchMEM benchmark represents a protein in a membrane surrounded by water, consisting of approximately 82,000 atoms, and was run with 10,000 steps and a 2 fs time step. The water benchmarks used water droplets with 3 different sizes: 5 nm, 10 nm, and 15 nm. These datasets contained 12,165, 98,319, and 325,995 atoms, respectively. Taken together, the various molecular structures and sizes in these 4 benchmarks allowed for a thorough evaluation of Gromacs performance under different simulation scenarios.

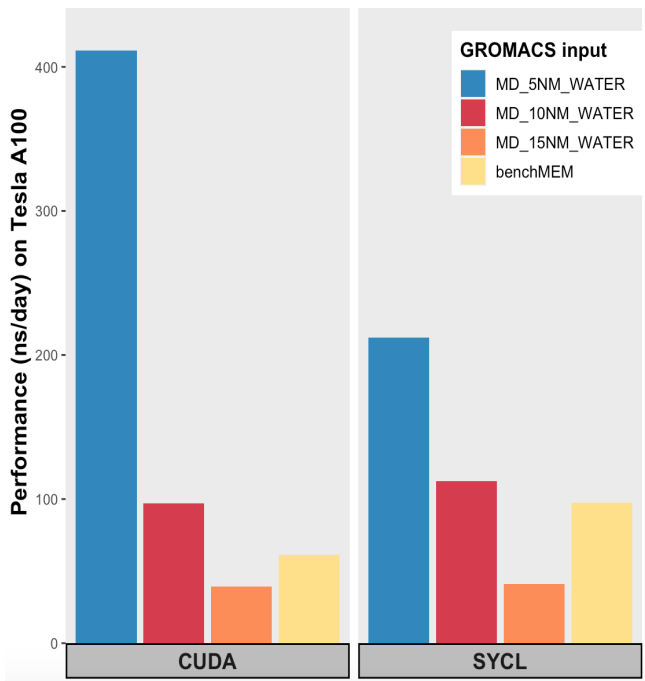

Fig. 1. Comparison of the performance of Gromacs CUDA (left) and Gromacs SYCL (right) for four different MD datasets run on a Nvidia Tesla A100 GPU. Larger values on the y-axis values indicate better performance.

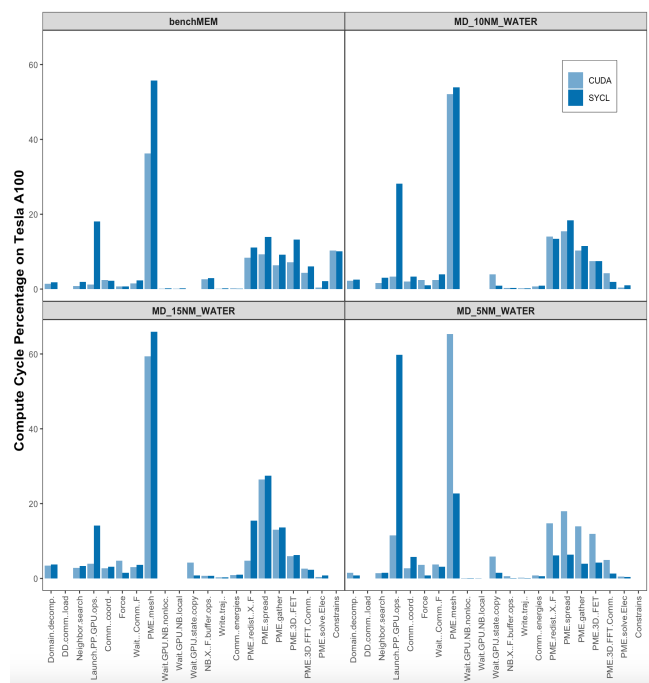

Fig. 2. Comparison of the percentage of cycles spent in each routine for Gromacs SYCL and Gromacs CUDA for four different MD datasets.

In Figure 1 we compare the performance of the CUDA and SYCL versions of Gromacs on an NVIDIA GPU system comprised of 8 A100s. The CUDA simulation showed much better performance than the SYCL simulation for the 5NM_WATER dataset. However, in the case of the constrained dataset benchMEM, 10 nm and 15 nm, SYCL performed better.

In Figure 2, the performance for the benchMEM and the three water datasets is shown as a function of the individual stages in the simulation for CUDA and SYCL, respectively. These figures provide a comprehensive view of the compute time spent in the various parts of the simulation and shows how the performance of the CUDA and SYCL simulations compare for each part. As can be seen in the figure, a significant portion of the simulation time is spent in the PME mesh routines which heavily rely on FFT (Fast Fourier Transform) computations. A possible explanation as to why SYCL outperformed CUDA for the larger datasets in Figure 1 is due to its ability to leverage hardware optimizations, including the utilization of Intel MKL library and SIMD instructions, which are specifically designed to improve accelerated FFT calculations.However, further studies are needed to truly understand the observed performance improvement of SYCL over CUDA in these larger datasets.

To assess the performance of Gromacs on different Nvidia GPU architectures, we conducted MD simulations using fixed numbers of cores and threads using P100, V100, and A100 GPUs. The performance results are presented in Figure 3. As expected, the A100 GPU exhibited superior performance for CUDA simulations, owing to its larger number of cores and other architectural advantages compared to the other GPUs. When comparing the performance of CUDA and SYCL, for the A100 GPU, SYCL outperformed CUDA specifically for the constrained benchMEM, 10 nm and 15 nm datasets.However, for the rest of the Nvidia GPUs, SYCL performed similarly to CUDA. The performance for the benchMEM and MD_15NM_WATER datasets is shown as a function of the individual stages in the simulation for CUDA and SYCL in Figures 4 and 5, respectively.

## V. CONCLUSION

The comparative performance analysis of SYCL and CUDA on Gromacs for different NVIDIA GPUs (V100, P100, and A100) using various water and protein datasets reveals that SYCL demonstrates competitive performance when compared to CUDA. The results of this study fall in line with previous comparisons of the two frameworks. In addition, this study suggests that SYCL can be a viable alternative to CUDA for molecular dynamics simulations using Gromacs, providing users with additional programming flexibility. SYCL's potential to improve cross-platform portability without compromising performance indicates it can be used to provide pathways for leveraging future advancements in computer architectures.

Future work could involve expanding the performance analysis to include other GPUs and datasets, as well as exploring the scalability of SYCL and CUDA for larger simulations. Such investigations would contribute to a deeper understanding of the capabilities and limitations of SYCL and CUDA for molecular dynamics simulations and guide researchers in selecting the most efficient and effective framework for their specific requirements.

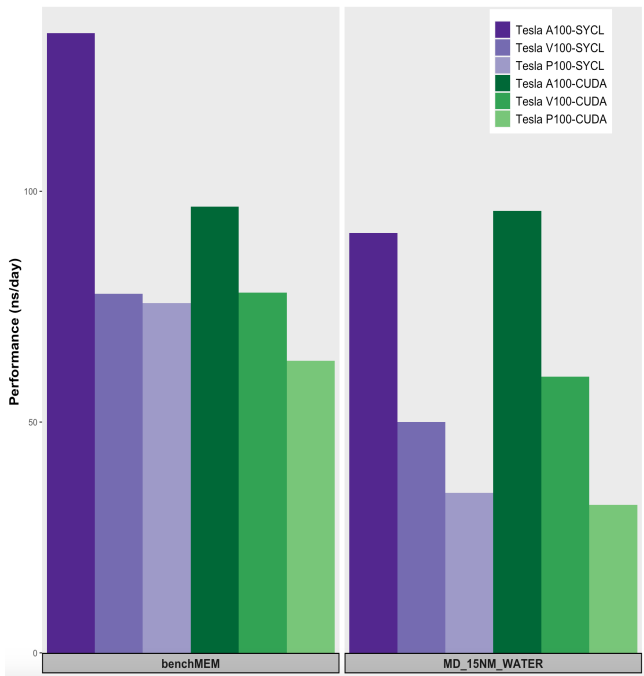

Fig. 3. Comparative performance analysis of Gromacs using SYCL and CUDA on P100,V100 and A100 for the benchMEM (left) and 15 nm water (right) datasets. Larger values on the y-axis values indicate better performance.

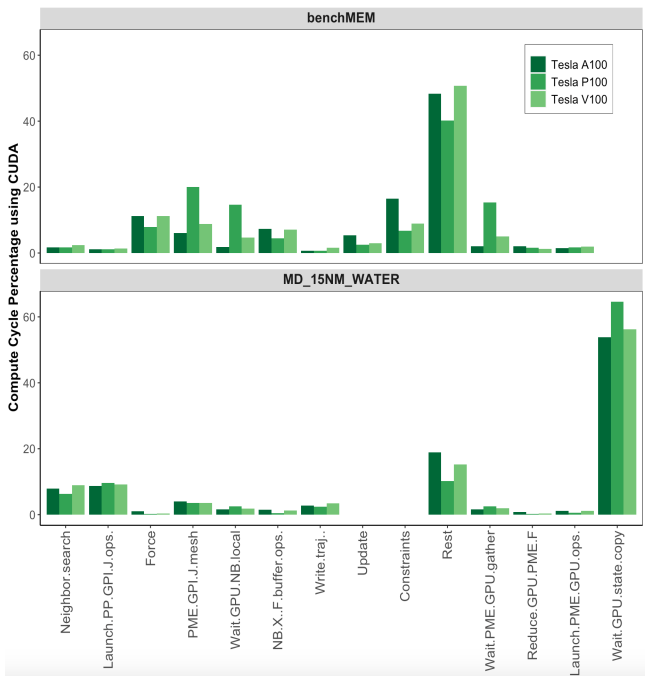

Fig. 4. Comparison of the percentage of cycles spent in each routine for Gromacs with CUDA using benchMEM (top) and 15 nm water (bottom) datasets on Tesla P100, V100 and A100.

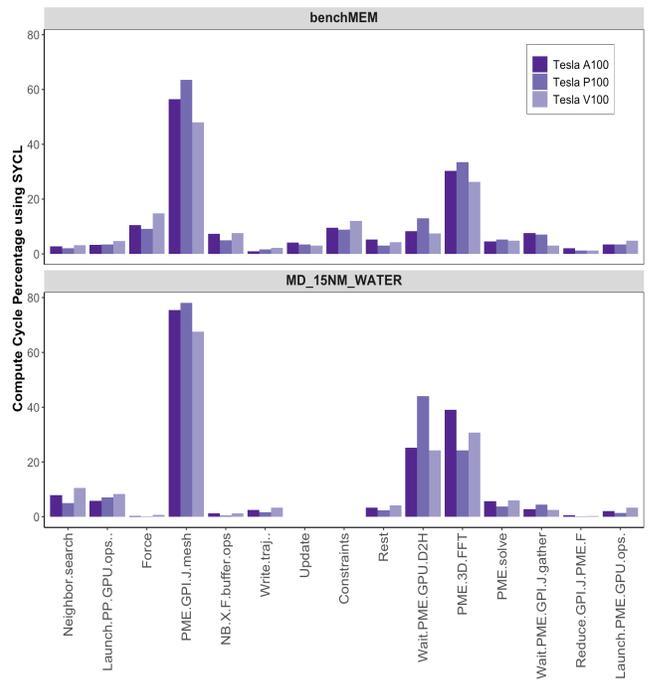

Fig. 5. Comparison of the percentage of cycles spent in each routine for Gromacs with SYCL for the benchMEM (top) and 15 nm water dataset (bottom) on Tesla P100, V100 and A100.

ACKNOWLEDGMENTS

We would like to thank Professor Huan-Xiang Zhou from the University of Illinois Chicago for his helpful advice and discussions.

We would also like to extend our thanks to the GROMACS support forum for their invaluable assistance in guiding us through the installation process of SYCL in our container environment. Their prompt and knowledgeable responses helped us overcome challenges and enabled us to harness the benefits of SYCL, enhancing the efficiency and performance of our simulations.

The A100 comparisons were run on machines at the Delta GPU cluster at NCSA through allocation TRA190023 from the Advanced Cyberinfrastructure Coordination Ecosystem: Services & Support (ACCESS) program, which is supported by National Science Foundation grants #2138259, #2138286, #2138307, #2137603, and #2138296.